\documentclass[sigconf]{acmart}

\AtBeginDocument{%
  }

\setcopyright{acmlicensed}
\copyrightyear{2026}
\acmYear{2026}
\acmDOI{XXXXXXX.XXXXXXX}
\acmConference[]{}{}{}
\acmISBN{978-1-4503-XXXX-X/2018/06}
\usepackage{subfigure}
\usepackage{subcaption}
\usepackage{multirow}

\begin{document}

\title{ArkTS-CodeSearch: A Open-Source ArkTS Dataset for Code Retrieval}

\author{Yulong~He}
\email{st065924@student.spbu.ru}
\orcid{0000-0002-3001-0096}
\affiliation{%
  \institution{St. Petersburg State University}
  \city{St. Petersburg}
  \country{Russia}
}
\author{Artem~Ermakov}
\email{irrationalkid201@icloud.com}
\orcid{0009-0004-1993-8135}
\affiliation{%
  \institution{ITMO University}
  \city{St. Petersburg}
  \country{Russia}
}
\author{Sergey~Kovalchuk}
\authornote{Corresponding author. Email: kovalchuk@itmo.ru}
\email{kovalchuk@itmo.ru}
\orcid{0000-0001-8828-4615}
\affiliation{%
  \institution{ITMO University}
  \city{St. Petersburg}
  \country{Russia}
}
\author{Artem~Aliev}
\email{artem.aliev@gmail.com}
\orcid{0000-0001-7984-4721}
\affiliation{%
  \institution{St. Petersburg State University}
  \city{St. Petersburg}
  \country{Russia}
}
\author{Dmitry~Shalymov}
\email{wd.shalymov@spbu.ru}
\orcid{0000-0002-8794-7306}
\affiliation{%
  \institution{St. Petersburg State University}
  \city{St. Petersburg}
  \country{Russia}
}

\renewcommand{\shortauthors}{He et al.}

\begin{abstract}
ArkTS is a core programming language in the OpenHarmony ecosystem, yet research on ArkTS code intelligence is hindered by the lack of public datasets and evaluation benchmarks. This paper presents a large-scale ArkTS dataset constructed from open-source repositories, targeting code retrieval and code evaluation tasks. We design a single-search task, where natural language comments are used to retrieve corresponding ArkTS functions. ArkTS repositories are crawled from GitHub and Gitee, and comment–function pairs are extracted using tree-sitter-arkts, followed by cross-platform deduplication and statistical analysis of ArkTS function types.  We further evaluate existing open-source code embedding models on the single-search task and perform fine-tuning using both ArkTS and TypeScript training datasets, resulting in a high-performing model for ArkTS code understanding. This work establishes the first systematic benchmark for ArkTS code retrieval. Both the dataset and our fine-tuned model are available at \url{https://huggingface.co/hreyulog/embedinggemma_arkts} and \url{https://huggingface.co/datasets/hreyulog/arkts-code-docstring}.
\end{abstract}




\maketitle

\section{Introduction}

OpenHarmony has emerged as a widely adopted open-source operating system for smart devices, and ArkTS serves as its core programming language for application development. As the OpenHarmony ecosystem continues to grow~\cite{10.1145/3720538}, the demand for intelligent code understanding tools—such as code search, code recommendation, code translation~\cite{10.1145/3691620.3695362,10.1145/3728941} and automated code analysis—has increased accordingly~\cite{chen2025arkanalyzerstaticanalysisframework,10.1145/3696630.3728565,liu2025frameworkawarecodegenerationapi}. However, research on ArkTS code intelligence remains limited due to the absence of publicly available datasets and standardized evaluation benchmarks.

In recent years, significant progress has been made in code semantic search, driven by large-scale datasets and pre-trained code embedding models. Benchmarks such as CodeSearchNet~\cite{husain2019codesearchnet} have enabled systematic evaluation of natural language–to–code retrieval for mainstream programming languages including Python, Java, and JavaScript, among others, along with subsequent datasets such as CodeXGLUE~\cite{lu2021codexgluemachinelearningbenchmark} and AdvTest, which further expand task coverage and evaluation settings.

To address this gap, we construct the first large-scale ArkTS dataset for code retrieval. Our dataset is collected from both GitHub\footnote{https://github.com/} and Gitee\footnote{https://gitee.com/}, covering a wide range of OpenHarmony-related repositories. Using \texttt{tree-sitter-arkts}\footnote{https://pypi.org/project/tree-sitter-arkts-open/}, we parse ArkTS source code into abstract syntax trees (AST) and extract function-level code units together with their associated natural language docstrings. The resulting dataset is processed and reformatted into HuggingFace-compatible structures, facilitating reproducibility and future research.

Based on this dataset, we define a single-search code retrieval task, where natural language docstrings are used to retrieve corresponding ArkTS functions. We adopt the CodeSearchNet framework as a baseline retrieval protocol and conduct a comprehensive evaluation of existing open-source code embedding models on the ArkTS code search task. Furthermore, to better capture ArkTS-specific semantics, we fine-tune pre-trained embedding models using ArkTS training data, achieving improved retrieval performance.

Our contributions can be summarized as follows:
\begin{itemize}
    \item We construct and release the first large-scale, publicly available ArkTS dataset for code retrieval, collected from GitHub and Gitee.
    \item We establish a standardized benchmark for ArkTS code search by defining a single-search task and adopting  CodeSearchNet style evaluation metrics.
    \item We evaluate existing open-source code embedding models on ArkTS and present a fine-tuned model that demonstrates strong performance on ArkTS code retrieval.
\end{itemize}

We release both the dataset and the fine-tuned model to support future research on ArkTS code intelligence and to encourage further exploration of intelligent development tools for the OpenHarmony ecosystem.






\section{Related Work}

\subsection{ArkTS Language}

ArkTS is the primary programming language of the OpenHarmony ecosystem, designed to support efficient cross-platform application development across distributed devices~\cite{li2023softwareengineeringopenharmonyresearch}. As a TypeScript-derived language that underpins Huawei’s HarmonyOS ecosystem, ArkTS plays a critical role in the emerging post-mobile application landscape, yet remains largely underexplored in code intelligence research. Although large-scale code corpora such as The Stack~\cite{kocetkov2022stack3tbpermissively} cover hundreds of programming languages, rapidly growing ecosystems like OpenHarmony are severely underrepresented. This imbalance creates a pronounced representation gap, limiting systematic evaluation of language models on ArkTS-specific constructs—such as declarative UI annotations, distributed application primitives, and cross-language bindings—that fundamentally distinguish ArkTS from its TypeScript predecessor.

This gap is particularly consequential given the rapid expansion of HarmonyOS. With deployment targets exceeding 800 million devices worldwide and an ongoing transition toward a fully independent application ecosystem (HarmonyOS NEXT), there is an urgent need for scalable, reproducible evaluation infrastructure to support AI-assisted software development. Recent efforts, including static analysis frameworks (e.g., ArkAnalyzer~\cite{chen2025arkanalyzerstaticanalysisframework}) and preliminary studies on large language model–based ArkTS code generation, highlight both the technical distinctiveness of ArkTS and the immaturity of existing tooling support. Prior work has explicitly identified the limited exposure of foundation models to ArkTS code as a major bottleneck for downstream tasks such as automated library migration and cross-ecosystem code generation~\cite{10.1145/3728941}. Empirical studies further report notable performance degradation when translating established Java or TypeScript programming patterns into ArkTS, even when advanced prompting strategies are employed~\cite{10764996}.

A publicly available, high-quality ArkTS dataset directly addresses these challenges by enabling systematic benchmarking and reproducible experimentation. As HarmonyOS NEXT continues its large-scale rollout, such infrastructure is essential for advancing research on cross-ecosystem migration, defect detection, and multilingual code representation learning—core problems in data mining for underrepresented programming languages. By establishing standardized benchmarks that capture the unique syntactic and semantic characteristics of ArkTS, this effort lowers the barrier to entry for studying language evolution from TypeScript to ArkTS and promotes more equitable progress in developer tooling beyond dominant, Western-centric software ecosystems.

\begin{figure*}[t]
\centering
\includegraphics[scale=0.05]{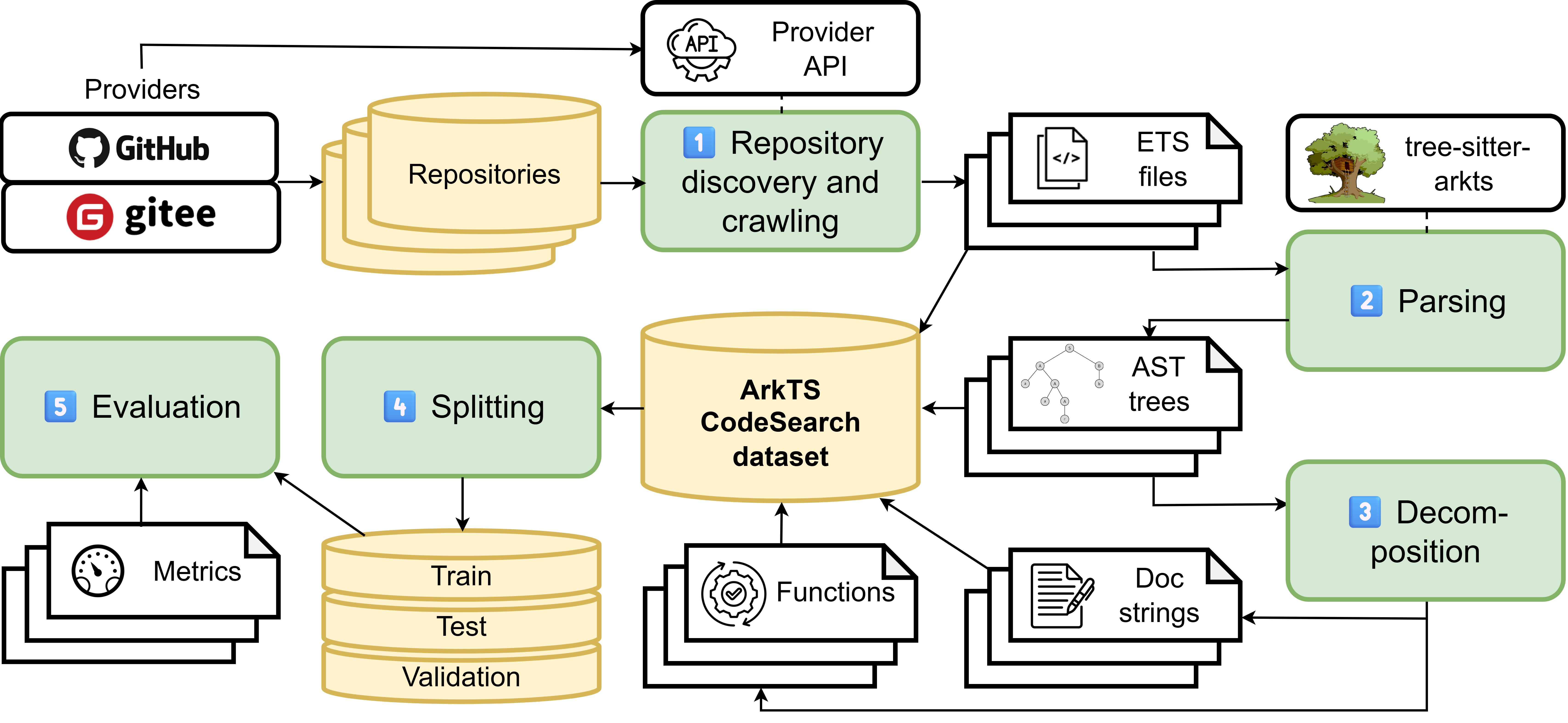}
\caption{Workflow for creating ArkTS-CodeSearch.} \label{pipeline}
\end{figure*}

\subsection{AST Parser}

AST parsers provide a structured and hierarchical representation of source code, explicitly encoding syntactic structure while implicitly capturing semantic relationships. AST-based representations have been widely adopted in code intelligence research, supporting tasks such as semantic code search, program analysis, and code summarization~\cite{10.1145/3212695}.

Compared with heuristic or regular-expression-based parsing, AST parsers offer improved robustness and syntactic correctness across diverse coding styles. Modern incremental parsers, such as Tree-sitter~\cite{tree-sitter}, enable scalable and language-aware code analysis and have been successfully applied to multiple programming languages. In this work, we leverage \texttt{tree-sitter-arkts} to parse ArkTS source files into ASTs, enabling accurate extraction of function- and class-level code units along with their associated docstrings. This structured preprocessing facilitates reliable code unit alignment and serves as a clean foundation for downstream code representation learning and evaluation.

\subsection{Code Semantic Search}

Code semantic search aims to retrieve relevant code snippets in response to a natural language query. Early approaches relied primarily on keyword matching or lexical similarity, which often fail to capture deeper semantic relationships between code and natural language. A commonly used method in this category is BM25\cite{10.1561/1500000019}, a sparse retrieval model based on term frequency–inverse document frequency (TF-IDF) scoring. 

Recent advances leverage neural models to learn joint embedding spaces for code and text, significantly improving retrieval performance. A range of pretrained models, such as CodeBERT~\cite{feng-etal-2020-codebert}, GraphCodeBERT~\cite{guo2021graphcodebertpretrainingcoderepresentations}, and CodeT5~\cite{wang2021codet5identifierawareunifiedpretrained}, have demonstrated strong performance on semantic code search tasks by exploiting large-scale multilingual and multi-language code corpora. Benchmarks such as CodeSearchNet have further standardized evaluation protocols and accelerated progress in this area. However, these models and benchmarks predominantly focus on widely used languages, leaving emerging languages such as ArkTS underexplored. Consequently, there is a lack of systematic evaluation of code embedding models on ArkTS, highlighting the need for dedicated datasets and benchmarks to assess semantic code search performance in this setting.

\section{Methodology}\label{sec3}

This section describes the overall methodology for constructing the ArkTS-CodeSearch dataset and training code retrieval models. 
Figure~\ref{pipeline} provides an overview of the complete workflow, including repository discovery, source code crawling, AST-based parsing and extraction, dataset construction, and model training and evaluation. 
Starting from raw ArkTS repositories collected from GitHub and Gitee, we apply structured preprocessing and filtering to obtain high-quality docstring--code pairs. 
The resulting dataset is then used to benchmark existing open-source embedding models and to perform domain-specific fine-tuning for ArkTS semantic code search.
\subsection{Data Collection}

To construct the ArkTS-CodeSearch dataset, we assembled a corpus of ArkTS source code from the two primary platforms hosting OpenHarmony-related projects: GitHub and Gitee. Our data collection pipeline comprises three main stages: (1) repository discovery, (2) source code crawling and filtering, and (3) structured code parsing and extraction, as detailed in the following subsections.

\subsubsection{Repository Discovery}

We employed two complementary strategies to identify ArkTS repositories, ensuring coverage from both the broader open-source community and the official OpenHarmony ecosystem.

\textit{GitHub}. We performed a systematic search on GitHub using the primary keyword \texttt{arkts}, which is a common identifier for projects related to OpenHarmony and ArkTS. This keyword-based discovery aims to capture a diverse range of repositories from independent developers and open-source contributors.

\textit{Gitee}. To include projects curated within the official HarmonyOS/ OpenHarmony environment, we sourced repositories directly from the dedicated OpenHarmony exploration page on Gitee\footnote{\url{https://gitee.com/explore/harmony}}. This platform-specific curated list provides high-quality, domain-relevant repositories that complement the GitHub search results.

The combination of broad keyword-based discovery (GitHub) and a focused, curated source (Gitee) is designed to mitigate platform-specific bias and construct a representative dataset of ArkTS projects.



\begin{figure*}[t]
\centering
\includegraphics[scale=0.35]{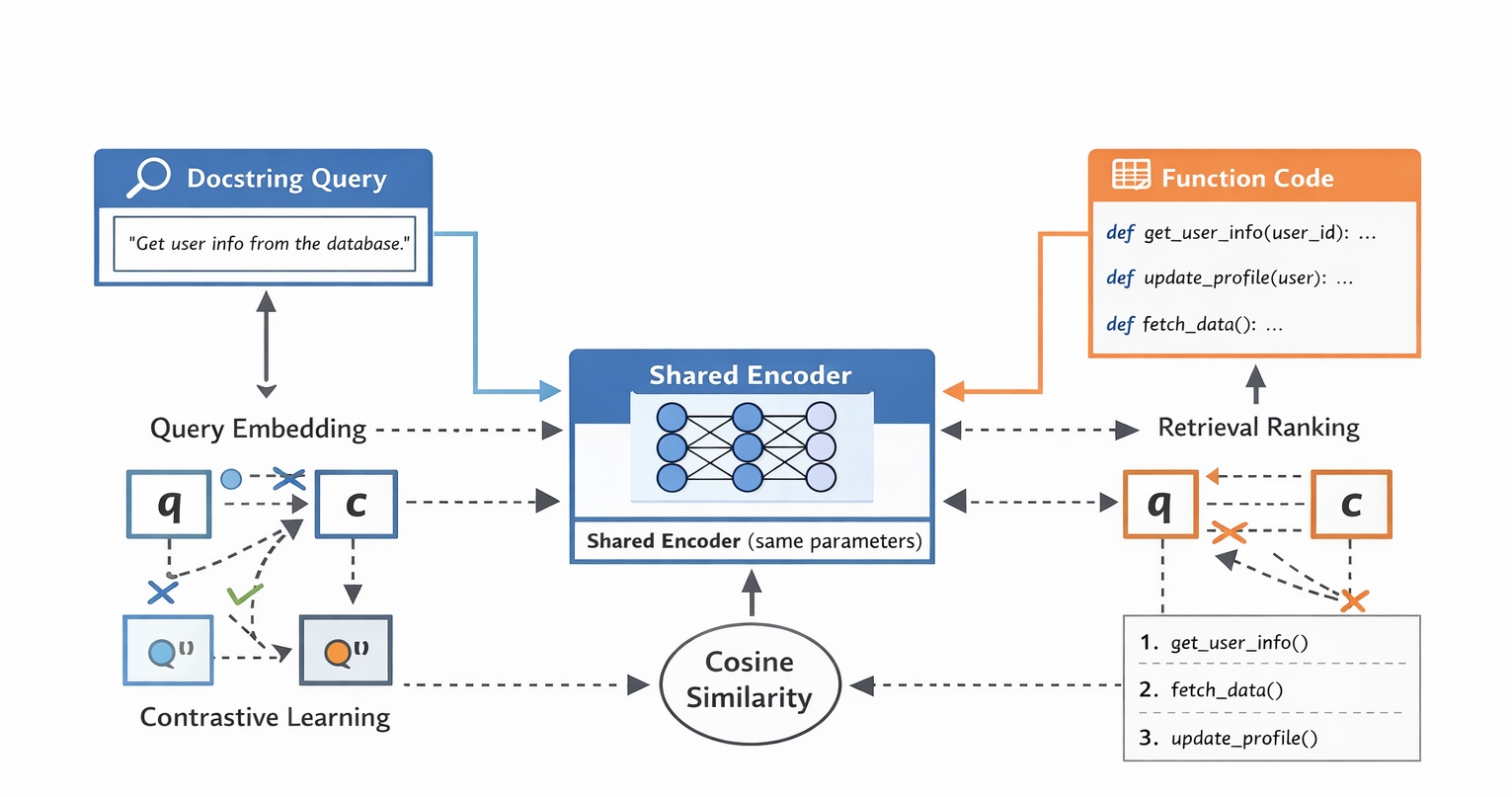}
\caption{Overview of the CodeSearchNet-style retrieval framework. 
Docstrings and function code are encoded by a shared encoder into a unified embedding space. 
The model is trained with contrastive learning and retrieves functions based on cosine similarity.} \label{codesearchnet}
\end{figure*}
\subsubsection{Repository Crawling}


At the initial stage of the search, lists of candidate repositories from both platforms were obtained. To ensure that there is no redundancy in the case, we performed the deduplication step. After removing the duplicates, our final unique set of repositories consisted of 1,577 repositories (560 from GitHub, 1,017 from Gitee).

For each of these repositories, we recursively looked through the directory structure to find the ArkTS source files. The files were filtered based on their extensions, saving files with the standard extension ArkTS (\texttt{.ets}) and discarding all other artifacts. For each valid ets file, we recorded the basic metadata, including the repository name, commit hash, file path, and source platform.

\subsubsection{Code Parsing and Extraction}

To extract structured code units, we employ \texttt{tree-sitter-arkts} to parse ArkTS source files into ASTs. Based on the AST representation, we extract function- and class-level code units along with their associated docstrings. Each extracted record consists of a natural language docstring, the corresponding ArkTS function or class implementation, and its AST-based representation.

The use of AST parsing ensures syntactic correctness and enables accurate extraction of code units across different coding styles. The resulting records are aggregated into a unified dataset, forming the foundation for downstream code retrieval and evaluation tasks.

\subsubsection{Licensing Considerations} The dataset was derived from open-source repositories hosted on GitHub and Gitee, which collectively encompass diverse license types including Apache, MIT, BSD, MPL, and others prevalent in the OpenHarmony ecosystem. The derived dataset -- comprising function-docstring pairs extracted via AST parsing -- constitutes a transformative derivative work intended for research purposes under fair use principles. However, users are advised to verify compliance with source repository licenses for any production or commercial deployment, as certain licenses may impose additional obligations (e.g., copyleft provisions).

\subsection{Retrieval Methods}

We evaluate two approaches for code retrieval. First, as a classic lexical retrieval method, BM25 ranks documents based on term frequency and inverse document frequency. For Chinese-heavy ArkTS docstrings, we segment both queries and code snippets using the \texttt{jieba tokenizer}~\footnote{https://github.com/fxsjy/jieba}, which combines dictionary-based matching with statistical modeling, constructs a directed acyclic graph of possible word sequences, and uses a Hidden Markov Model to handle new words. Despite its simplicity, BM25 provides a strong sparse baseline for comparison.

Second, semantic search leverages dense vector representations of queries and code snippets to capture semantic relationships beyond lexical overlap. Given a query $q$ and a code snippet $c$ , both are encoded into a shared embedding space using a pretrained code model, and their similarity is measured by cosine similarity \cite{reimers2019sentence}:

$$
\text{sim}(q, c) = \frac{\mathbf{v}_q \cdot \mathbf{v}_c}{\|\mathbf{v}_q\| \|\mathbf{v}_c\|}.
$$

Code snippets are ranked according to similarity scores, enabling the retrieval of semantically relevant results even without direct lexical overlap, which is particularly important for ArkTS with bilingual or multilingual documentation.

We adopt the CodeSearchNet paradigm for ArkTS semantic code retrieval, formulating the task as docstring-to-function retrieval. Each data instance consists of a function-level code snippet paired with its corresponding docstring. As shown in Figure~\ref{codesearchnet}, a shared encoder is used for both docstrings and code, and the model is trained with contrastive learning to align semantically corresponding pairs in the embedding space. Retrieval is then performed via cosine similarity.

\subsection{Evaluation Metrics}
To assess retrieval performance, we adopt widely used metrics for retrieval-based code search. In our experiments, each query is searched against the entire test set to evaluate retrieval effectiveness.
\subsubsection{MRR}
Mean Reciprocal Rank (MRR) measures the average reciprocal rank of the first relevant code snippet for each query~\cite{Craswell2009}. It reflects how early the correct function is retrieved and is defined as:
\[
\text{MRR} = \frac{1}{|Q|} \sum_{i=1}^{|Q|} \frac{1}{\text{rank}_i},
\]
where \(\text{rank}_i\) denotes the rank position of the first relevant result for query \(i\), and \(|Q|\) is the total number of queries.

\subsubsection{NDCG}
Normalized Discounted Cumulative Gain (NDCG)~\cite{10.5555/1394399} accounts for the positions of relevant code snippets in the ranked list:
\[
\text{NDCG}@k = \frac{DCG@k}{IDCG@k}, \quad 
DCG@k = \sum_{i=1}^{k} \frac{2^{rel_i}-1}{\log_2(i+1)},
\]
where \(rel_i\) denotes the relevance score of the result at position \(i\), and \(IDCG@k\) is the ideal DCG for normalization.

\subsubsection{Recall}
Recall@k measures the proportion of queries for which the relevant code snippet appears within the top \(k\) retrieved results:
\[
\text{Recall@k} = \frac{1}{|Q|} \sum_{i=1}^{|Q|} \mathbb{I}(\text{rank}_i \le k),
\]
where \(\mathbb{I}(\cdot)\) is the indicator function.

\subsection{Fine-Tuning}

To enhance the semantic alignment between natural language descriptions and code functions, we fine-tune pre-trained code embedding models using a supervised contrastive learning objective, with the goal of improving retrieval performance on the ArkTS code search task.

\subsubsection{Training Data Construction}
For fine-tuning, we primarily use our ArkTS dataset, which provides function–docstring pairs suitable for supervised contrastive learning. 
In addition, we incorporate a large-scale TypeScript corpus introduced by \citet{lozhkov2024starcoder}, from which we extract 340,116 function–docstring pairs using tree-sitter-typescript parsing. All extracted pairs are processed using the same bidirectional alignment strategy as in ArkTS to ensure consistency.

We choose TypeScript as an auxiliary training source because ArkTS is a language derived from TypeScript, sharing similar syntax, type systems, and programming abstractions. 
As a result, TypeScript code provides semantically and structurally related supervision that can serve as an effective intermediate domain for model adaptation.

Based on these data sources, we explore three training configurations:

(1) \textbf{ARKTS-TRAINING}, which fine-tunes the model using only ArkTS pairs;

(2) \textbf{TS-TRAINING}, which fine-tunes the model using only the TypeScript-derived pairs; and

(3) \textbf{TS$\rightarrow$ARKTS-TRAINING}, a two-stage strategy that first adapts the model on TypeScript data and then further fine-tunes it on ArkTS.

These configurations allow us to systematically study the impact of related-language pre-adaptation and domain-specific fine-tuning on ArkTS code retrieval performance.

\subsubsection{Model Architecture and Input Processing}

We adopt the \texttt{Senten\allowbreak ceTransformer} framework~\cite{reimers2019sentence} to load and fine-tune pre-trained code embedding models. 
Docstrings and function bodies are encoded into dense vector representations using a shared encoder, ensuring that both modalities are mapped into the same embedding space. 
The maximum input length is set to 512 tokens, and the embedding dimensionality follows the configuration of the underlying pre-trained model.

\subsubsection{Training Objective}
Fine-tuning is conducted using the Multiple Negatives Ranking Loss~\cite{reimers-2019-sentence-bert}, a contrastive learning objective widely adopted in dense retrieval. Within each batch, positive docstring--function pairs are contrasted against other examples in the batch, which are treated as implicit negatives. This encourages the model to assign higher similarity scores to true pairs while distancing unrelated ones. Conceptually, this can be viewed as a form of in-batch contrastive learning~\cite{oord2019representationlearningcontrastivepredictive}, which brings semantically aligned docstring--function pairs closer in the embedding space and pushes mismatched pairs further apart. Such a training strategy has been shown to be highly effective in semantic retrieval tasks, resulting in more discriminative representations and improved code retrieval performance.

Formally, given a query embedding \(q\), a positive code embedding \(c^{+}\), and a set of negative code embeddings \(\{c^{-}_i\}_{i=1}^{N}\), the loss is defined as:
\[
\mathcal{L} = - \log \frac{\exp(\mathrm{sim}(q, c^{+}))}{\exp(\mathrm{sim}(q, c^{+})) + \sum_{i=1}^{N} \exp(\mathrm{sim}(q, c^{-}_i))},
\]
where \(\mathrm{sim}(\cdot)\) denotes cosine similarity. This in-batch contrastive objective encourages semantically aligned pairs to cluster in the embedding space while separating mismatched pairs~\cite{oord2019representationlearningcontrastivepredictive}.

\subsubsection{Optimization Details}
All models are trained for two epochs with a batch size of 32 and a learning rate of $2\times10^{-5}$, using the Adam optimizer with a linear warm-up schedule for 10\% of total steps. Training samples are shuffled each epoch to improve generalization.

\section{Dataset Statistics}\label{sec5}

We provide a comprehensive statistical analysis of our ArkTS dataset to better understand its characteristics. 
The dataset was collected on January 29, 2026, and is publicly available on HuggingFace\footnote{Available at \url{https://huggingface.co/datasets/hreyulog/arkts-code-docstring}}.

In total, the dataset contains 24,452 ArkTS functions extracted from open-source repositories, forming paired docstring–function samples for code retrieval tasks.
Considering that ArkTS is not exclusively developed on GitHub and that a large number of Chinese developers actively contribute on Gitee, we collect data from both platforms to achieve broader and more representative coverage.

\begin{figure}[t]
\centering
\includegraphics[scale=0.3]{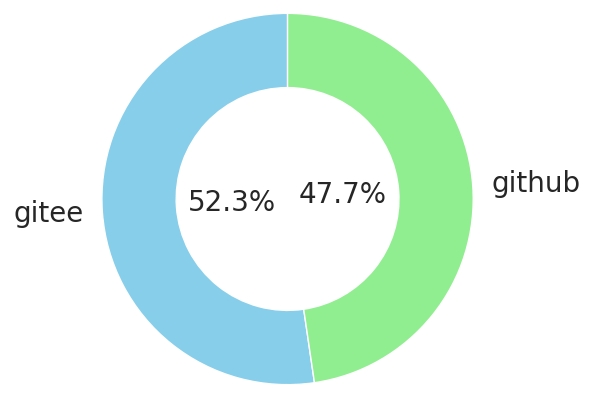}
\caption{Distribution of source }\label{source}
\end{figure}

Specifically, 12,792 functions are collected from Gitee and 11,660 from GitHub, resulting in a balanced cross-platform distribution. Figure~\ref{source} illustrates the distribution of source platforms.

By incorporating repositories from both GitHub and Gitee, our dataset captures diverse development practices, coding styles, and project structures across different developer communities.
\begin{figure}[t]
\centering
\includegraphics[scale=0.45]{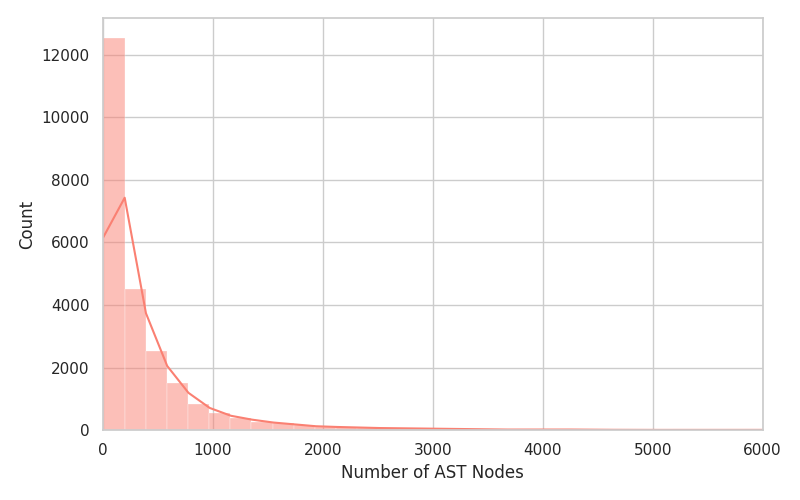}
\caption{Distribution of AST Length (\# of nodes) }\label{ast}
\end{figure}

\begin{figure}[t]
\centering
\includegraphics[scale=0.45]{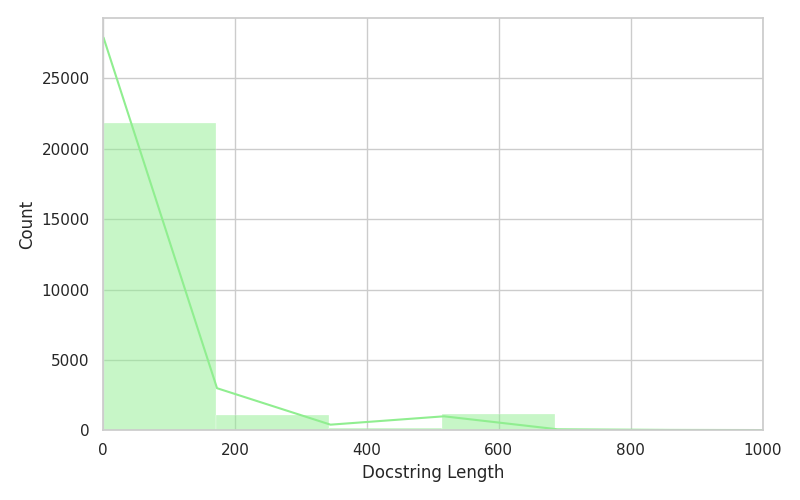}
\caption{Distribution of Docstring Length (characters)} \label{docstring}
\end{figure}
Figure~\ref{ast} shows the distribution of AST lengths measured by the number of nodes in each function. Most functions have relatively small ASTs, while a few outliers exhibit significantly larger structures, indicating complex code snippets.

Figure~\ref{docstring} illustrates the distribution of docstring lengths measured in characters. 
While the majority of functions are accompanied by short docstrings (primarily within 0--200 characters), 
the distribution exhibits a noticeable secondary peak around 500 characters. 
The presence of this secondary peak indicates heterogeneous documentation patterns within ArkTS projects, where most functions are lightly documented, while a smaller subset receives more comprehensive descriptions, likely corresponding to core logic or publicly exposed interfaces. Overall, the distribution remains right-skewed with a long tail, 
indicating substantial variability in documentation practices across projects.

\begin{figure}[t]
\centering
\begin{subfigure}{}
     \includegraphics[width=0.45\textwidth]{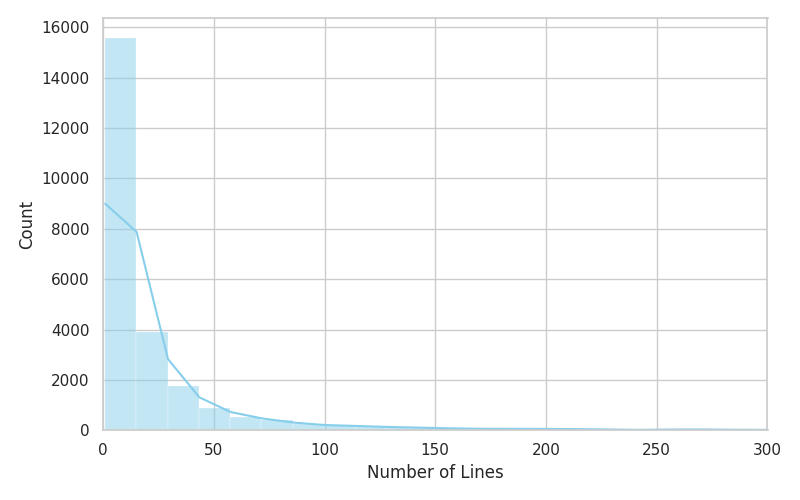}
     
\end{subfigure}\begin{subfigure}{}
     \includegraphics[width=0.45\textwidth]{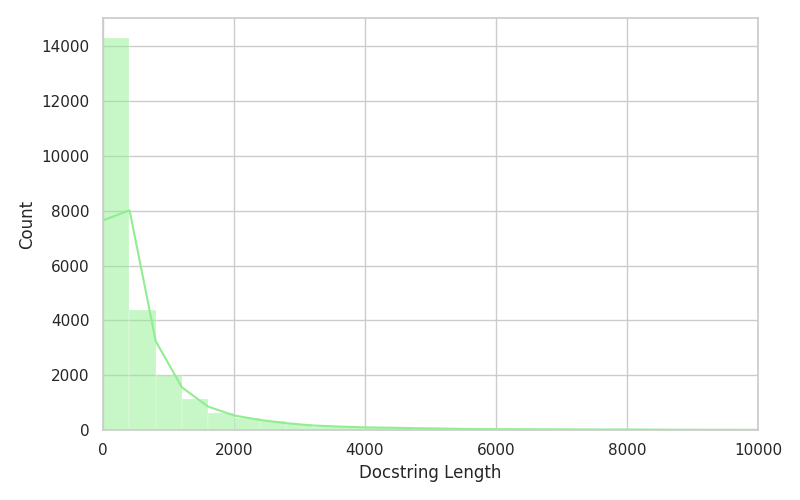}
     
\end{subfigure}

\caption{Distribution of Function Length in ArkTS Dataset}
\label{function_length}
\end{figure}

Figures~\ref{function_length} depict the distributions of function lengths in characters and lines, respectively. Similar to docstrings, function lengths exhibit a skewed distribution, with most functions being relatively short but a few reaching substantial sizes. These statistics highlight the diversity in code complexity within the dataset.

For experimental evaluation, the dataset is randomly split into training, validation, and test sets following an 8:1:1 ratio, resulting in 19,561 training samples, 2,445 validation samples, and 2,446 test samples.

Overall, these statistics demonstrate that our ArkTS dataset is diverse in terms of function complexity, documentation, and source repositories, making it suitable for training and evaluating code retrieval models.

\section{Experiments and Results}
In this section, we evaluate the effectiveness of different embedding models on the ArkTS code retrieval task.
We report results on the held-out test set using standard information retrieval metrics, including Mean Reciprocal Rank (MRR), NDCG@5, and Recall@K.
All models are evaluated under the same retrieval setting to ensure fair comparison.
The evaluation is conducted using our open-source retrieval evaluation framework,\footnote{Available at \url{https://github.com/hreyulog/retrieval_eval}.}
\begin{table*}[]
\centering
\begin{tabular}{l|l|llll|llll}
\hline
\multirow{2}{*}{Models} & \multirow{2}{*}{Params}  & \multicolumn{4}{c|}{Original} & \multicolumn{4}{c}{Finetuned (ARKTS-TRAINING)} \\ 
 & & MRR & NDCG@5 & Recall@1 & Recall@5 & MRR & NDCG & Recall@1 & Recall@5 \\ \hline
\textbf{google/embeddinggemma-300m }& 308M & 0.6399 & 0.6654 & 0.5740 & 0.7416 &\textbf{ 0.7788} &\textbf{ 0.8034} &\textbf{ 0.7142 }&\textbf{ 0.8769} \\
QWEN/QWEN3-EMBEDDING-0.6B & 596M & \textbf{0.6776} & \textbf{0.7015} &\textbf{ 0.6141} &\textbf{ 0.7723} & 0.7745 & 0.7971 & 0.7146 & 0.8643 \\
BAAI/bge-m3 & 567M & 0.5283 & 0.5603 & 0.4464 & 0.6558 & 0.7467 & 0.7729 & 0.6750 & 0.8504 \\
BAAI/bge-base-zh-v1.5 & 110M & 0.3598 & 0.3903 & 0.2841 & 0.4816 & 0.6748 & 0.7055 & 0.5961 & 0.7972 \\
BAAI/bge-base-en-v1.5 & 110M & 0.3439 & 0.3637 & 0.2935 & 0.4227 & 0.4866 & 0.5144 & 0.4195 & 0.5977 \\
intfloat/e5-base-v2 & 110M & 0.3073 & 0.3261 & 0.2596 & 0.3823 & 0.4855 & 0.5135 & 0.4158 & 0.5973 \\
BM25 (jieba tokenizer) & - & 0.2043 & 0.2204 & 0.1643 & 0.2690 & - & - & - & - \\ 
\hline
\end{tabular}
\caption{Comparison of original open-source models and finetuned models (ARKTS-TRAINING)}
\label{tab:merged_leaderboard}
\end{table*}

\begin{table*}[htbp]
\centering
\begin{tabular}{l|l|llll|llll}
\hline
\multirow{2}{*}{Models} & \multirow{2}{*}{Params} & \multicolumn{4}{c|}{TS-TRAINING} & \multicolumn{4}{c}{TS-ARKTS-TRAINING} \\ 
 & & MRR & NDCG & Recall@1 & Recall@5 & MRR & NDCG & Recall@1 & Recall@5 \\ \hline
\textbf{google/embeddinggemma-300m} & 308M &0.6259& 0.6525 & 0.5572 & 0.7318 & \textbf{0.8013} & \textbf{0.8229} & \textbf{0.7412} & \textbf{0.8867} \\
Qwen/Qwen3-Embedding-0.6B & 596M & \textbf{0.6522} &\textbf{0.6785} &\textbf{ 0.5809} & \textbf{0.7563} & 0.7708 & 0.7947 & 0.7065 & 0.8655 \\
BAAI/bge-m3 & 567M & 0.6051 & 0.6338 & 0.5311 & 0.7191 & 0.7581 & 0.7846 & 0.6872 & 0.8630 \\
BAAI/bge-base-zh-v1.5 & 110M & 0.5270 & 0.5571 & 0.4509 & 0.6468 & 0.6759 & 0.7032 & 0.6018 & 0.7841 \\
intfloat/e5-base-v2 & 110M & 0.3751 & 0.3955 & 0.3226 & 0.4567 & 0.4557 & 0.4821 & 0.3904 & 0.5613 \\
BAAI/bge-base-en-v1.5 & 110M & 0.3834 & 0.4047 & 0.3299 & 0.4681 & 0.4539 & 0.4770 & 0.3949 & 0.5457 \\
\hline
\end{tabular}
\caption{Leaderboard for finetuned models under different training datasets (TS-TRAINING vs TS-ARKTS-TRAINING)}
\label{tab:op_leaderboard_combined}
\end{table*}





\subsection{Leaderboard for Opensorce Models}

Table~\ref{tab:merged_leaderboard} reports the performance of several widely used open-source embedding models evaluated on the ArkTS code retrieval task without any task-specific fine-tuning.
In addition, we include a traditional lexical retrieval baseline, BM25 with the jieba tokenizer, for comparison.

These models differ substantially in parameter scale, pretraining objectives, and language coverage.

Among all evaluated models, \texttt{Qwen/Qwen3-Embedding-0.6B}~\cite{qwen3embedding} achieves the best overall performance, with an MRR of 0.6776 and Recall@5 of 0.7723.
This indicates that large-scale general-purpose embedding models pretrained on multilingual and code-related corpora are more effective at capturing the semantic relationship between ArkTS docstrings and code.

The \texttt{google/embeddinggemma-300m}~\cite{embedding_gemma_2025} model ranks second across all metrics, despite having roughly half the number of parameters compared to Qwen3.
This suggests that model architecture and pretraining data quality play an important role beyond sheer model size.

The multilingual \texttt{BAAI/bge-m3}~\cite{bge-m3} model achieves competitive performance and ranks third overall.

Focusing on models with approximately 100M parameters, \texttt{BAAI/\allowbreak bge-base-zh-v1.5}~\cite{bge_embedding} consistently outperforms its English counterpart (\texttt{bge-base-en-v1.5}) and \texttt{intfloat/e5-base-v2}~\cite{wang2022text} across all evaluation metrics.
A plausible explanation is that a considerable portion of ArkTS docstrings in our dataset are written in Chinese. As a result, Chinese-oriented embedding models are better aligned with the linguistic characteristics of the ArkTS documentation.

In contrast, general-purpose sentence embedding models primarily trained on natural language data, such as \texttt{e5-base-v2}~\cite{wang2022text}, exhibit noticeably lower performance.

We also observe that the BM25 baseline with jieba tokenization performs substantially worse than all embedding-based methods, achieving an MRR of 0.2043, Recall@1 of 0.2204, Recall@5 of 0.1643, and Recall@10 of 0.2690.
This result suggests that purely lexical matching is insufficient for ArkTS code retrieval, where semantic alignment between docstrings and code snippets is crucial.

This performance gap highlights the domain and language mismatch between generic text embedding models and the requirements of code semantic search, particularly for programming languages like ArkTS that combine source code with bilingual or multilingual documentation.

Overall, these results demonstrate that directly applying off-the-shelf embedding models to ArkTS code retrieval yields limited performance, thereby motivating the need for domain- and language-aware optimization.

\subsection{Fine-Tuning}

\begin{figure*}[t]
  \centering
  \includegraphics[scale=0.65]{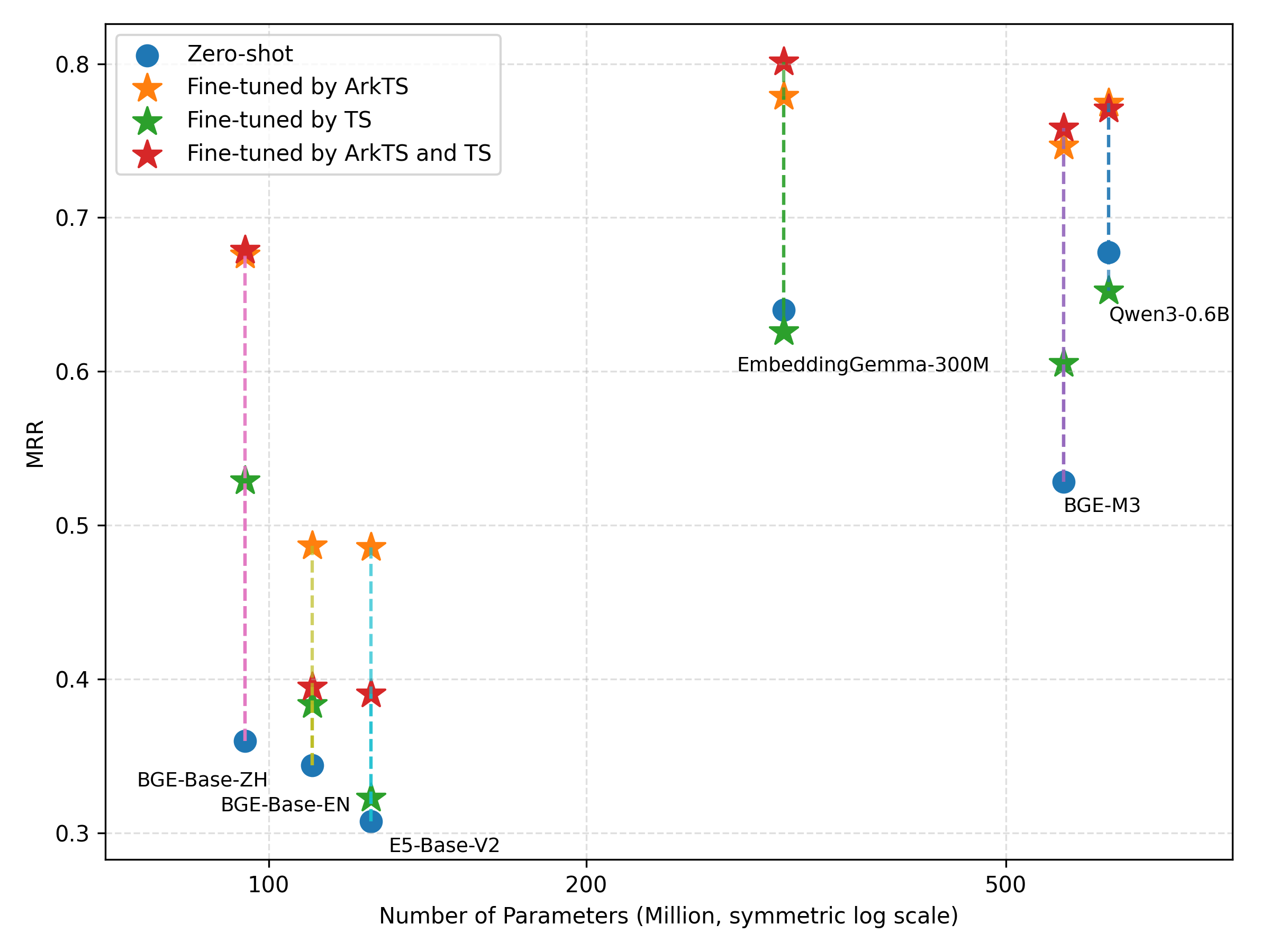}
  \caption{Model size versus MRR before and after fine-tuning on the ArkTS dataset.
A small horizontal jitter is applied for clarity of visualization.}
  \label{performance_size}
\end{figure*}

Fine-tuning substantially improves retrieval performance across all embedding models compared to zero-shot baselines, demonstrating the effectiveness of task-specific supervision for aligning natural-language docstrings with ArkTS code. Evaluation results are summarized in Tables~\ref{tab:merged_leaderboard} and~\ref{tab:op_leaderboard_combined}, and a well-performing fine-tuned model trained on the ArkTS training set is publicly released on Hugging Face.\footnote{Available at \url{https://huggingface.co/hreyulog/embedinggemma_arkts}} The best-performing model is available upon reasonable request.

The choice of training data has a pronounced impact on retrieval quality. Models adapted to ArkTS-specific supervision achieve consistently higher scores than those trained only on the TypeScript corpus, across all metrics including MRR, NDCG, Recall@1, and Recall@5. This highlights the critical role of domain-aligned fine-tuning in semantic code retrieval.

The proposed TS→ArkTS training strategy improves performance over single-stage fine-tuning in most cases, especially for models with larger parameter sizes.
The two-stage procedure benefits from the broader semantic coverage of TypeScript data followed by ArkTS-specific adaptation, leading to strong gains in early-recall metrics.
In contrast, models with approximately 100M parameters do not consistently benefit from this strategy and may experience performance degradation, potentially due to limited model capacity under successive fine-tuning stages.

Among all evaluated models, \texttt{google/embeddinggemma-300m} achieves the best performance under all three training configurations, reaching an MRR of 0.6259 with TS-TRAINING and 0.8013 with TS→ARKTS-TRAINING. Notably, this mid-sized model outperforms larger alternatives such as \texttt{Qwen/Qwen3-Embedding-0.6B}, indicating that effective domain adaptation can outweigh advantages from increased model scale. Nevertheless, \texttt{Qwen/Qwen3-Embedding\allowbreak -0.6B} exhibits substantial absolute improvements after ArkTS-specific fine-tuning, demonstrating that large pretrained models also benefit significantly from targeted supervision.

Multilingual and Chinese-oriented models, including \texttt{BAAI/bge-m3} and \texttt{BAAI/bge-base-zh-v1.5}, show strong and stable gains across all training settings, particularly in Recall@1, reflecting improved precision for top-ranked results. In contrast, English-centric sentence embedding models such as \texttt{BAAI/bge-base-en-v1.5} and \texttt{intfloat/e5-base-v2} remain less competitive.  We attribute this difference to the linguistic characteristics of the ArkTS corpus: a substantial portion of code comments and docstrings are written in Chinese. As a result, models pretrained with stronger Chinese language representations are better aligned with the natural-language queries and documentation present in the dataset, leading to improved semantic matching.

Overall, these results demonstrate that training data composition is as critical as model architecture for semantic code search. Supervised contrastive fine-tuning on ArkTS-specific data substantially reshapes model rankings, and the proposed TS→ARKTS two-stage strategy provides a practical and effective approach for adapting general-purpose embedding models to specialized programming languages.

     



Figure~\ref{performance_size} illustrates the relationship between model size and retrieval performance.
While larger models generally achieve stronger zero-shot results, fine-tuning on the ArkTS dataset
leads to substantial performance gains across all model scales. Notably, several medium-sized models
achieve performance comparable to or exceeding larger models after fine-tuning, highlighting the
importance of domain-specific data adaptation.

\section{Conclusion and future work}

Our experiments and dataset analysis provide several key insights into ArkTS code semantic search and representation learning. The ArkTS dataset exhibits substantial diversity in both code structure and documentation practices: while most functions are short and lightly documented, long-tailed distributions in AST size, function length, and docstring length indicate that a non-negligible subset of the codebase involves complex logic and detailed explanations. The prevalence of Chinese-language docstrings further highlights the importance of language-aware modeling, as Chinese-oriented embeddings consistently outperform English-oriented counterparts in retrieval tasks.

Zero-shot evaluation demonstrates the limitations of directly applying off-the-shelf embedding models to ArkTS. Fine-tuning on ArkTS-specific data consistently yields significant gains, reshaping model rankings: mid-sized models can surpass larger pretrained models, indicating that domain-specific supervision and task-aligned contrastive learning are more critical than model scale alone. Leveraging TypeScript data in a two-stage TS$\rightarrow$ARKTS fine-tuning strategy further improves performance, suggesting that pre-adaptation on related languages is a practical approach for emerging programming languages.

Looking forward, several directions are promising. Incorporating commit messages, diffs, and additional OpenHarmony projects could extend the dataset to support commit-based code retrieval, code change understanding, and software evolution analysis. Structure-aware models that explicitly leverage AST or graph-based representations may further enhance semantic alignment between code and natural language. Expanding multilingual annotations would also facilitate broader cross-language and cross-platform code intelligence research. We hope this dataset and benchmark will serve as a foundation for future research on ArkTS and contribute to the development of more effective code intelligence techniques within the OpenHarmony ecosystem.

\bibliographystyle{ACM-Reference-Format}
\bibliography{custom1}










\end{document}